\documentclass[namedreferences]{SolarPhysics}
\usepackage{spr-sola-addons} 
\usepackage{color}           
\usepackage{url}             

\usepackage[pdftex]{graphicx} 
\usepackage{hyperref}

\newcommand{\degr}{{\hbox{$^\circ$}}}

\newcommand{\arcsec}{{\hbox{$^{\prime\prime}$}}}
\renewcommand{\sun}{_\odot}

\newcommand{\be}{\begin{equation}}
\newcommand{\ee}{\end{equation}}
\newcommand{\bea}{\begin{eqnarray}}
\newcommand{\eea}{\end{eqnarray}}
\newcommand{\beas}{\begin{eqnarray*}}
\newcommand{\eeas}{\end{eqnarray*}}

\newcommand{\eg}{{\it e.g.}}

\newcommand{\cf}{{\it c.f.}}
\renewcommand{\tabcolsep}{0.18cm}

\chardef\us=`\_ 

\begin{document}

\begin{article}

\begin{opening}

\title{Measurement of the Height of the Chromospheric Network Emission from {\it Solar Dynamics Observatory} images}

\author[addressref={aff1},corref,email={calissan@cc.uoi.gr}]{\inits{C.E.}\fnm{C.E.}~\lnm{Alissandrakis}}

\address[id=aff1]{Section of Astro-Geophysics, Department of Physics, University of Ioannina, GR-45110 Ioannina, Greece}

\runningauthor{C. E. Alissandrakis}
\runningtitle{Height of network emission}

\begin{abstract}
We measured the height of the chromospheric network in the 1700, 1600, and 304\,\AA\ wavelength bands of the {\it Atmospheric Imaging Assembly} (AIA) onboard the {\it Solar Dynamics Observatory} (SDO) from the shift of features on the disk with respect to corresponding features in SDO/{\it Helioseismic and Magnetic Imager} (HMI) images of the absolute value of the longitudinal magnetic field. We found that near the limb the 304\,\AA\ network emission forms $3.60\pm0.24$ Mm above the 1600\,\AA\ emission, which, in turn, forms $0.48\pm0.10$\,Mm above the HMI (6173 \AA) level. At the center of the disk the corresponding height differences are $2.90\pm0.02$\,Mm and $0.39\pm0.06$\,Mm respectively. We also found that the 1600\,\AA\ network emission forms $0.25\pm0.02$\,Mm above the 1700\,\AA\ emission near the limb and $0.20\pm0.02$\,Mm at the disk center. Finally, we examined possible variations with the solar cycle. Our results can help to check and refine atmospheric models.
\end{abstract}
\end{opening}

\section{Introduction} \label{intro} 
The height of formation of electromagnetic radiation in the solar atmosphere is determined by the local physical conditions and the transfer of the radiation. In classic empirical atmospheric models the height is not a basic parameter, but it is derived from the optical depth and the absorption coefficient (see, \eg, \citealp{1966soat.book.....Z}). The height at a particular frequency can be computed from the atmospheric model parameters (see, \eg. Figure 1 of \citealp{1976ApJS...30....1V}; Figure 2 of \citealp{2008ApJS..175..229A}). More recently and more relevant to the work presented here, \cite{2005ApJ...625..556F} computed the formation height of the {\it Transition Region and Coronal Explorer} (TRACE) wavelength bands at 1600 and 1700\,\AA; they gave values of 0.36 and 0.43\,Mm, with widths of 0.325 and 0.185\,Mm respectively. 

The observable quantity most directly associated with the height of formation is the solar radius. Measurements of the solar radius were compiled by \cite{2015ApJ...812...91R} for the optical and EUV and by \cite{2017SoPh..292..195M} for the microwave range (see also \citealp{2017ApJ...843..123B} for a measurement in X-rays). In a recent work (\citealp{2019SoPh..294...96A}; see also \citealp{2019SoPh..294..146A} for a correction), we measured the height of the limb with respect to the photosphere from TRACE and {\it Atmospheric Imaging Assembly} (AIA) images, while \cite{2018SoPh..293...20A} provided values of the limb height for a number of spectral lines and continua observed with the {\it Interface Region Imaging Spectrograph} (IRIS). As the region of formation of the radiation has a certain vertical extent, the limb height represents the upper bound of this region rather than the average; moreover, chromospheric spicules will further increase the measured height. These factors may explain in part why our measured values of 1.2\,Mm from TRACE at 1600\,\AA\ and 0.7 and 1.1\,Mm at 1700 and 1600\,\AA\ from AIA are above those computed by \cite{2005ApJ...625..556F}.

Indirect measurements of formation-height differences above sunspots, based on the time delay of 3-minute oscillations, were reported by \cite{2015ARep...59..959D}, who estimated that the 304\,\AA\ emission forms no more than 0.8\,Mm above the 1600\,\AA\ emission; \cite{Pats} used the time delay of $p$-mode waves between 100\,GHz images and 1600\,\AA\ images to estimate a height difference of about 1.2\,Mm.

On the solar disk, height differences translate to shifts between features observed at different wavelengths. Although \cite{2009A&A...507L..29F} measured the height difference between the 630\,nm Fe{\sc i} photospheric lines, no direct measurements of formation heights above the photosphere have been reported so far. Such measurements require high spatial resolution and accurate centering of images at different wavelengths. In a recent work \citep{2018A&A...619L...6N} on observations of the chromospheric network with the {\it Atacama Large mm and sub-mm Array} (ALMA), we reported that the 304\,\AA\ emission forms 3 Mm above the 1600\,\AA\ emission, both observed with the AIA onboard the {\it Solar Dynamics Observatory} (SDO). This conclusion was reached by comparing the position of the network structures imaged in the two AIA wavebands for the seven fields of view, from disk center to the limb, observed  with ALMA. That result could not be attributed to errors in pixel size, since the solar radius we measured at 304\,\AA\ was within 0.6\arcsec\ of the average value measured by \cite{2007A&A...476..369G} with the {\it Extreme Ultraviolet Imager Telescope} (EIT). Moreover, \cite{2019SoPh..294...96A}, using the 2012 Venus transit, compared the scale and orientation of the AIA band images to those of the {\it Helioseismic and Magnetic Imager} (HMI) images and demonstrated that deviations from the values in the headers of the image files were negligible.

In this work we expand our data set and we compute the height of the network emission for the 1700, 1600, and 304\,\AA\ AIA bands with respect to the continuum level observed by HMI at 6173\,\AA. In the second section we describe our observations, in Section 3 we present our results, and in Section 4 we discuss our conclusions.

\begin{figure}
\begin{center}
\includegraphics[width=\hsize]{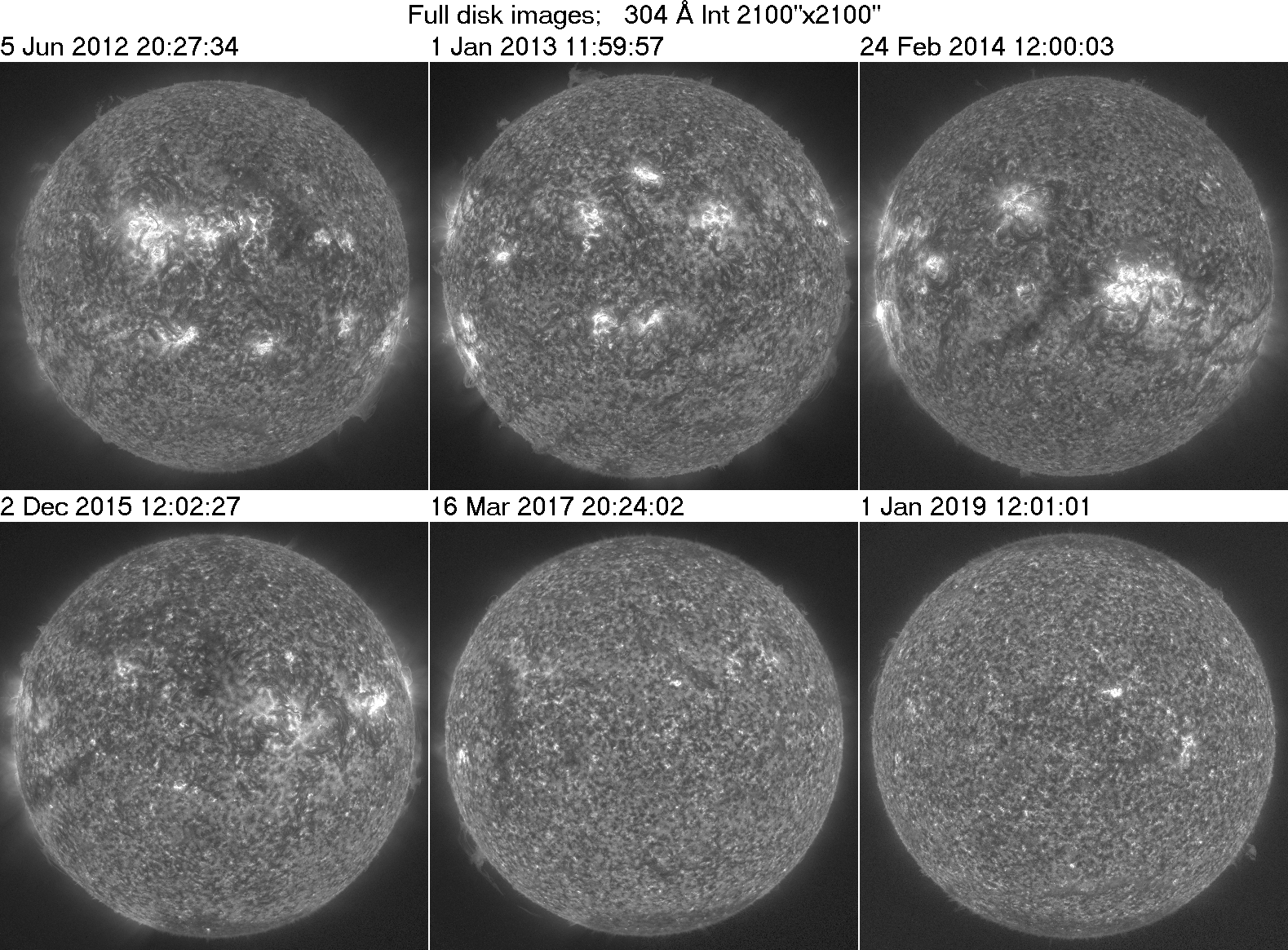}
\end{center}
\caption{Full disk images in the AIA 304\,\AA\ band selected for our analysis.}
\label{Fig:FD}
\end{figure}

\section{Observations and Data Reduction} \label{obs}
For the present study we used the AIA \citep{2012SoPh..275...17L} and HMI \citep{2012SoPh..275..229S} data set that we had employed in conjunction with the ALMA observations \citep{2018A&A...619L...6N}, as well as those that we had used for our works with IRIS data \citep{2018SoPh..293...20A}
and the Venus transit \citep{2019SoPh..294...96A}. In order to check for solar cycle effects, we expanded our data set by including observations from 2013, 2015, and 2019; we thus covered a period from 2012 to now (2019), that is from before the solar maximum up to the Solar Cycle 24\,--\,25 minimum of activity (see Figure \ref{Fig:FD}, where we give full-disk images for the selected days in the 304\,\AA\ band). For each data set we averaged a five-minute time series of images in order to reduce the effect of $p$-mode oscillations which are quite prominent in the 1600 and 1700\,\AA\ bands, and in order to reduce noise in the AIA EUV wavebands. 
The pixel size is 0.5\arcsec\ for HMI and $\approx0.6$\arcsec\ for AIA, while the spatial resolution is of the order of 1\arcsec.

We applied the scale corrections with respect to HMI computed by \cite{2019SoPh..294...96A} to the AIA images; although these corrections are small, they amount to $\approx0.15$\,Mm at the limb and can have non-negligible effects on our measurements. Moreover, we centered the HMI continuum and the AIA 1600 and 1700\,\AA\ images by performing a fit of the limb to a circle; the resulting corrections were small, of the order of 0.5 pixel.

For the measurement of the shift between images at different bands, we divided each image into 12 sectors, 30\degr\ wide; each sector was further divided into a number of zones, 60\arcsec\ wide, from just inside the limb down to a distance of about 0.25\,R$\sun$ from the center of the disk. Images near the limb were corrected for center-to-limb effects, which can be very different at different wavelengths; this was done by subtracting the average value of the intensity at a given distance from the limb. Subsequently, the image segments were remapped to polar coordinates and the radial shift between two wavelength bands was measured by cross correlation. 
Typical values of the maximum cross correlation are $\approx0.9$ between 1600 and 1700\,\AA\ images and $\approx0.4$ between 304 and 1600\,\AA\ images; we note that, even in the latter case, the peak stands clearly above the background. 
We add that in all wavelength bands network structures are optically thick, and this facilitates the comparison.

We thus obtained the shift as a function of position angle (measured counterclockwise from the North) and radial distance from the center of the Sun. The average shift at each location was computed after dropping the highest and the lowest shift value; this accounts for excess shifts due to the presence of sunspots, filaments {\it etc\/}. in some images. Finally, from the average shift, the corresponding height difference was computed as a function of radial distance This method is better than the comparison of selected individual regions as was done by \cite{2018A&A...619L...6N}, because it gives a global picture for the solar disk and can be used to check both for Pole--Equator differences and for the change of height between the disk center and the limb.

\begin{figure}
\begin{center}
\includegraphics[width=.48\hsize]{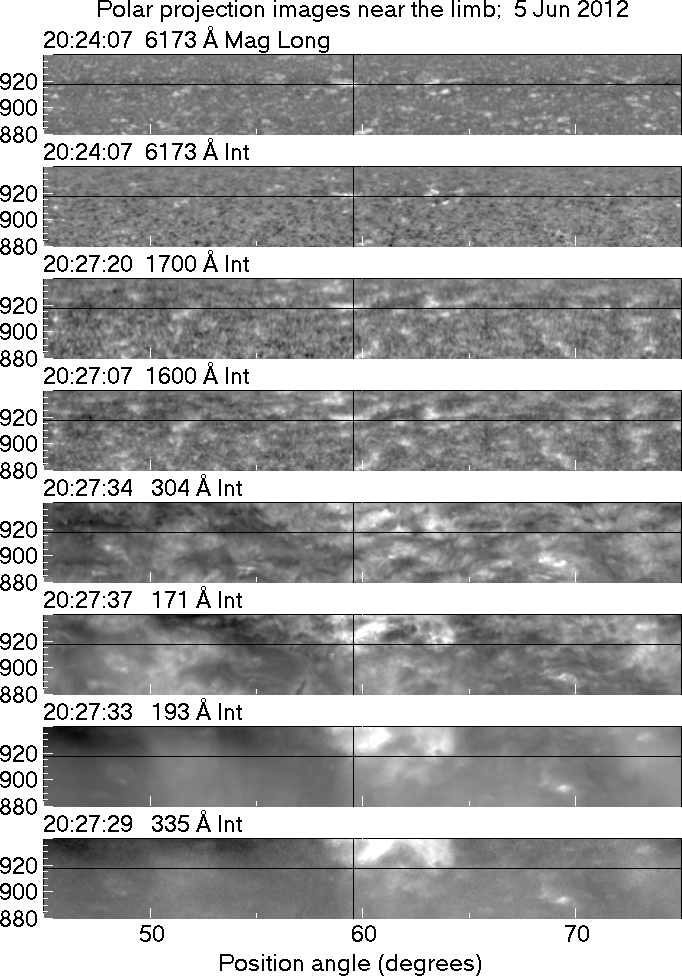}~~\includegraphics[width=.48\hsize]{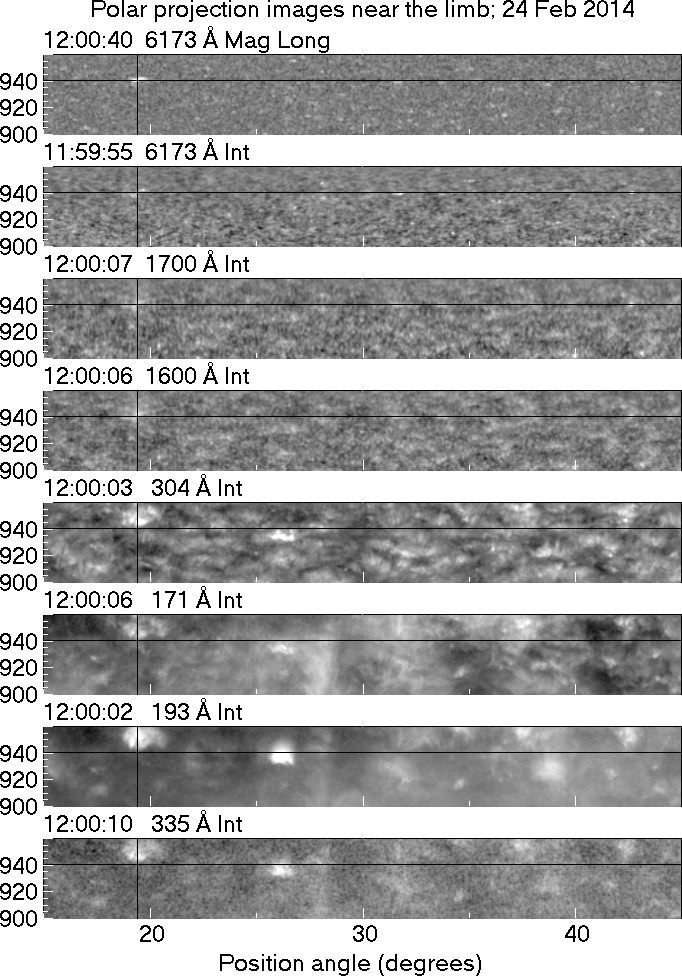}
\end{center}
\caption{Polar-projection of a region near the limb, HMI and AIA images; the {\it top panel} shows the absolute value of the longitudinal magnetic field. The {\it vertical coordinate} is the distance from the center of the disk in arc sec. The {\it cross-hair} marks the position of prominent magnetic features.}
\label{Fig:polar}
\end{figure}

As a proxy for the network structures in HMI we used the absolute value of the longitudinal magnetic field, $|B_l|$; to improve the quality of the image, we performed a smoothing with a 1\arcsec\ Gaussian. We computed the shifts of network features in 1700, 1600, and 304\,\AA\ with respect to HMI, of 1600\,\AA\ with respect to 1700\,\AA, and of 304\,\AA\ with respect to 1600\,\AA; for the latter we smoothed the 1600\,\AA\ image by a 3\arcsec\ Gaussian, in order to compensate for the increase of the size of the network features between the two wavelengths.

\section{Results}\label{res}
Figure \ref{Fig:polar} shows polar-projection images for two regions; in addition to HMI, 1700, 1600 and 304\,\AA, we also provide 171, 193 and 335\,\AA\ images for reference. It is obvious in the figure that the features marked by the cross-hair, as well as many others, are shifted towards the limb in the 304\,\AA\ band. 

We note that network features are clearly visible in the HMI, 1700, 1600 and 304\,\AA\ bands; they are still visible in the 171\,\AA\ band, but this is apparently due to the fact that the temperature response of this channel extends to temperatures lower than its nominal temperature of $\approx10^6$\,K \citep{2014SoPh..289.2377B}. No network features are seen in the 193 and 335\,\AA\ bands, in agreement with the well-known fact that the network disappears in the corona \citep{1974ApJ...188L..27R}.

\begin{figure}[h]
\begin{center}
\includegraphics[width=\hsize]{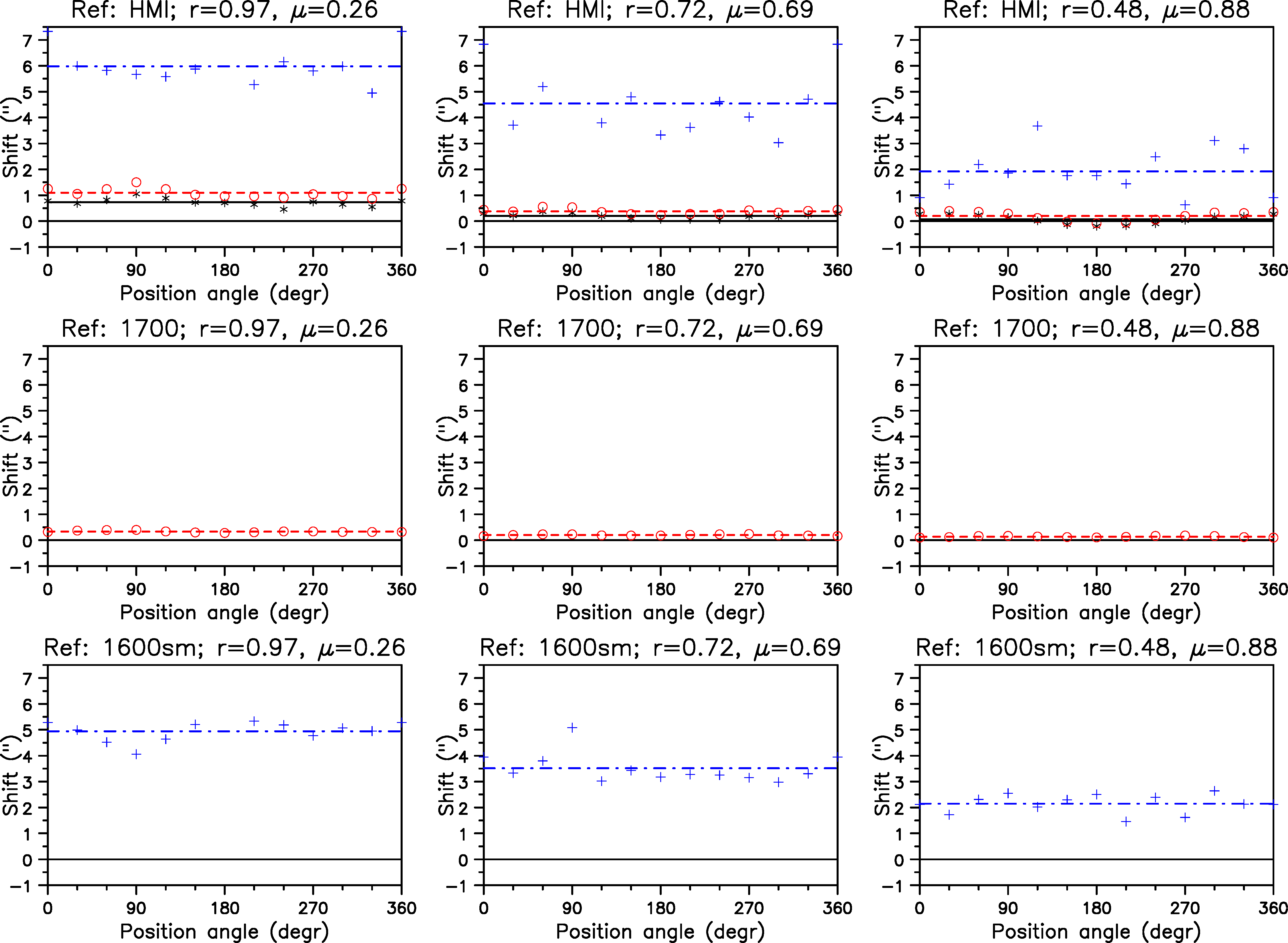}
\end{center}
\caption{The radial image shift with respect to a reference image as a function of position angle for three locations on the solar disk for the data set of 18 April, 2017. The reference image, the distance from the center of the disk in solar radii and the corresponding cosine of the heliocentric angle, $\mu$, are marked above each panel. {\it In the last row} the 3\arcsec\ smoothed 1600\,\AA\ image was used as a reference. The {\it horizontal lines} mark the shift averaged over the position angle. {\it Black asterisks and full lines} are for 1700\,\AA, {\it red, circles and dotted lines} for 1600\,\AA\ and {\it blue, crosses and dash-dotted lines} for 304\,\AA.}
\label{Fig:shift_pa}
\end{figure}

An example of our measurements of the image shift as a function of position and distance from the disk center is given in Figure \ref{Fig:shift_pa} for the 2017 data, which is typical of all of our data sets. We note that in all cases the shift decreases towards the center of the disk, as expected. We can see no Pole--Equator differences, not even when coronal holes were present near the Poles, as in the 2015, 2017, and 2019 data sets. 

Fluctuations around the mean value provide an estimate of the accuracy of our measurements. For the same wavelength, their variation with position angle is not the same at all disk locations, which excludes pointing errors as their origin; they can be due to intrinsic structural differences and/or image distortion, which is different in different bands. There are no measurable fluctuations when the 1600 and 1700\,\AA\ bands are compared (middle plot in Figure \ref{Fig:shift_pa}), which are imaged by the same telescope and have very similar structure. We note further that the fluctuations of the 304\,\AA\ band are smaller when it is compared to the smoothed 1600\,\AA\ band image than when it is compared with HMI, apparently due to the larger structural differences in the latter case.

\begin{figure}
\begin{center}
\includegraphics[width=\hsize]{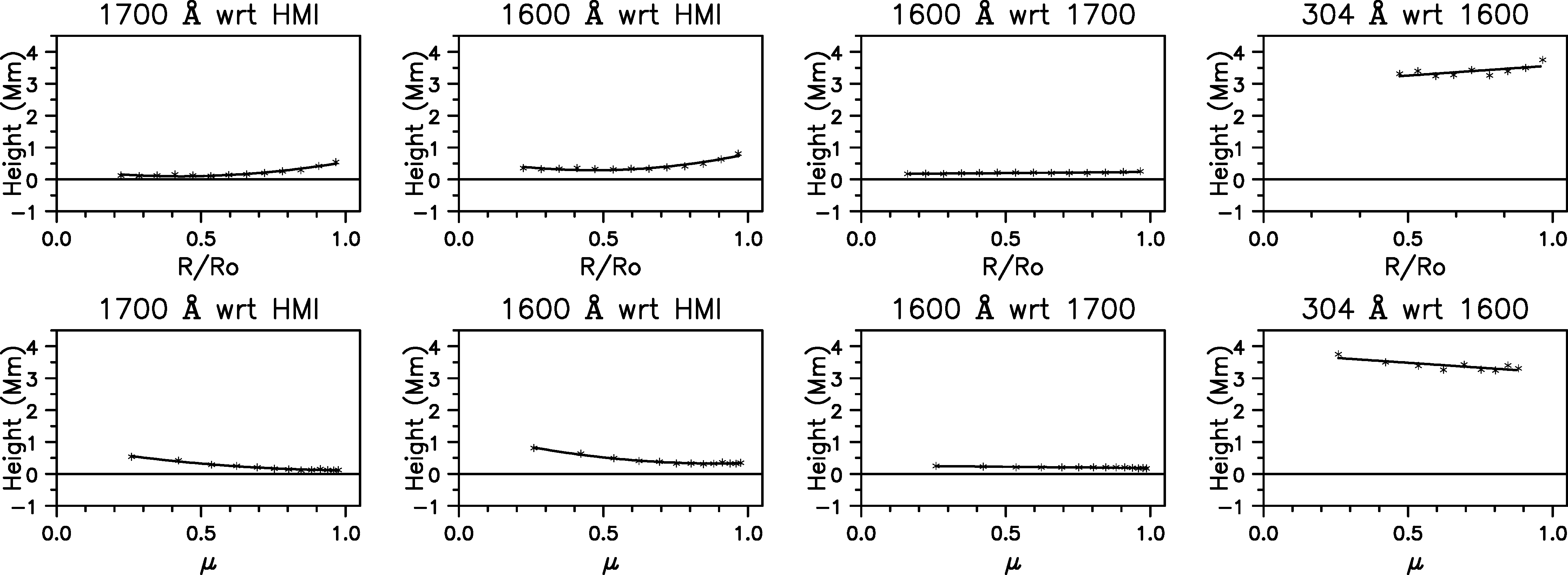}
\end{center}
\caption{The height of network emission, plotted as a function of distance from the center of the disk {\it (top row)} and as a function of the cosine of the heliocentric angle, $\mu$, for the data set of 18 April, 2017. The reference wavelength and the image wavelength are marked above each panel. {\it Solid lines} show the result of linear regression for 304\,\AA\ and a quadratic regression for the other wavelengths.}
\label{Fig:height_r}
\end{figure}

Figure \ref{Fig:height_r} shows the emission height, deduced from the average shift, as a function of distance from the center of the disk (top row) and as a function of the cosine of the heliocentric angle ($\mu$, bottom row), for the same data set shown in Figure \ref{Fig:shift_pa}. It is quite clear that the emission height increases from the center of the disk to the limb. For the 1700 and 1600\,\AA\ bands this increase is better represented by a second degree curve, both as a function of $R$/R$\sun$ and as a function of $\mu$, while for the 304\,\AA\ band the scatter of values does not permit anything else but a linear fit.

\begin{figure}[h]
\begin{center}
\includegraphics[width=\hsize]{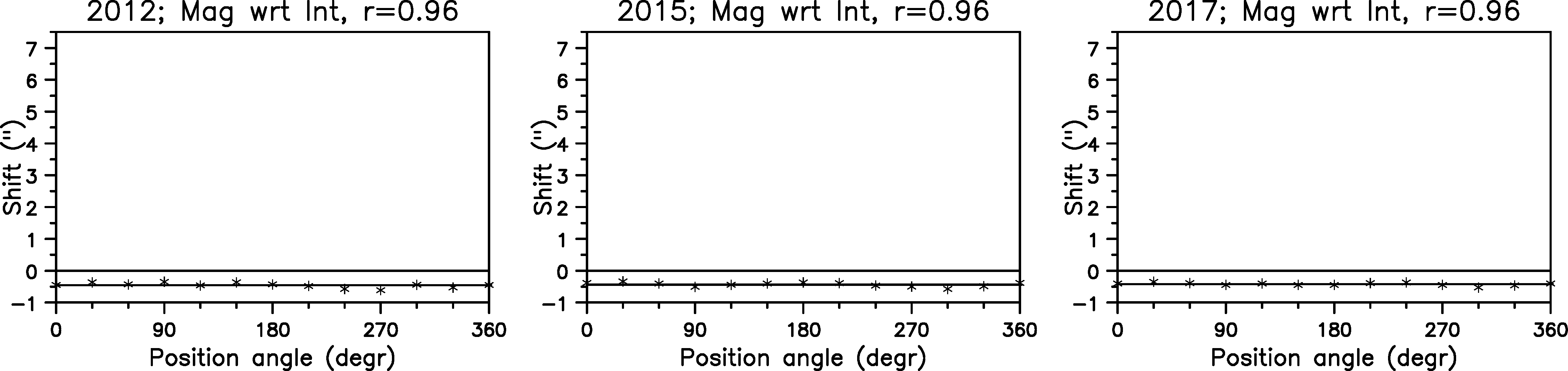}
\end{center}
\caption{The radial image shift near the limb of HMI continuum images with respect to images of the absolute value of the magnetic field, as a function of position angle for three data sets.}
\label{Fig:height_mag_int}
\end{figure}

It turned out that $|B_l|$ is not a perfect proxy for the network at the level of the HMI continuum. Indeed, a comparison of $|B_l|$ images with continuum intensity images near the limb, where network structures have enough contrast to be detectable (\cf\ Figure \ref{Fig:polar}), showed that the apparent height of the former is slightly lower (Figure \ref{Fig:height_mag_int}). 

The very small fluctuations of the shift, with root mean square (RMS) values from 0.03 to 0.06\arcsec, indicate that this is a real effect. Averaged over all six data sets, the height difference is $0.34\pm0.07$\,Mm. A possible explanation of this effect is that the magnetic field is nearly radial at the center of a structure and inclined at its edge; consequently, for an optically thick structure very close to the limb, the peak of the longitudinal magnetic field will not coincide with the maximum field intensity, which is what the network brightness approximately represents, but it will be shifted towards the disk center (a well-known effect with sunspots). Far from the limb, this effect should be negligible.

Considering all of the issues discussed above, we adopted the following procedure to estimate the emission height:
\begin{itemize}
\item Correct the height near the limb, measured for 1700 and 1600\,\AA\ with respect to $|B_l|$, for the shift reported in the previous paragraph.
\item Estimate the height of the 304\,\AA\ emission with respect to the HMI level by adding the height difference between 1600\,\AA\ and HMI to the height of 304\,\AA\ with respect to 1600\,\AA.
\item In all cases the height at the disk center was estimated by extrapolation of the height {\it vs.} $\mu$ curves to $\mu=1$.
\end{itemize}

\begin{table}[h]
\caption{Formation heights [Mm] of the network near the solar limb}
\label{Table:heightsLimb}
\begin{tabular}{lrrrrrr}
\hline
Date        & $S_n$ &\multicolumn{1}{c}{1700-HMI}&\multicolumn{1}{c}{1600-HMI}&\multicolumn{1}{c}{1600-1700}&\multicolumn{1}{c}{304-1600}&\multicolumn{1}{c}{304-HMI}\\
\hline
5 Jun. 2012  &  86.6 & $0.23\pm0.10$ & $0.53\pm0.12$ & $0.25\pm0.07$ & $3.45\pm0.38$ & $3.98\pm0.50$ \\
1 Jan. 2013  &  86.8 & $0.19\pm0.13$ & $0.47\pm0.13$ & $0.28\pm0.03$ & $3.41\pm0.31$ & $3.88\pm0.44$ \\
24 Feb. 2014 & 110.5 &$-0.03\pm0.17$ & $0.29\pm0.29$ & $0.26\pm0.03$ & $3.30\pm0.39$ & $3.59\pm0.68$ \\
2 Dec. 2015  &  57.8 & $0.20\pm0.10$ & $0.47\pm0.09$ & $0.26\pm0.02$ & $3.67\pm0.47$ & $4.14\pm0.56$ \\
16 Mar. 2017 &  25.7 & $0.20\pm0.07$ & $0.47\pm0.10$ & $0.25\pm0.02$ & $3.75\pm0.20$ & $4.22\pm0.30$ \\
1 Jan. 2019  &   5.4 & $0.42\pm0.17$ & $0.64\pm0.19$ & $0.22\pm0.02$ & $4.02\pm0.71$ & $4.66\pm0.90$ \\
\hline
Average     &       & $0.20\pm0.13$ & $0.48\pm0.10$ & $0.25\pm0.02$ & $3.60\pm0.24$ & $4.07\pm0.89$ \\
\hline
Limb height&       & 0.7~~~~~~~~~~~&1.1~~~~~~~~~~~&  &&5.4~~(peak)\\
\hline
\end{tabular}
\end{table}

The results near the limb are presented in Table \ref{Table:heightsLimb}, where we have added the 13-month smoothed International Sunspot Number, $S_n$;\footnote{http://sidc.be/silso/} we also give the height of the limb for 1700 and 1600\,\AA, as well as the height of peak intensity for 304\,\AA\ from \cite{2019SoPh..294...96A}. Error estimates are from the RMS of the shift values averaged over position angle, as described above. Results for the center of the disk are given in Table \ref{Table:heightsCD}; here the errors were estimated from RMS of the least squares  fit since, as mentioned previously, these values were computed by extrapolation of the height to $\mu=1$.

\begin{table}
\caption{Formation heights [Mm] of the network at the center of the solar disk}
\label{Table:heightsCD}
\begin{tabular}{lrccccc}
\hline
Date        & $S_n$ &   1700-HMI    &   1600-HMI    &   1600-1700   &   304-1600    &   304-HMI     \\
\hline
5 Jun. 2012  &  86.6 & $0.18\pm0.05$ & $0.41\pm0.03$ & $0.21\pm0.01$ & $2.75\pm0.19$ & $3.16\pm0.22$ \\
1 Jan. 2013  &  86.8 & $0.10\pm0.07$ & $0.38\pm0.03$ & $0.22\pm0.01$ & $2.96\pm0.13$ & $3.34\pm0.16$ \\
24 Feb. 2014 & 110.5 & $0.12\pm0.03$ & $0.33\pm0.04$ & $0.20\pm0.01$ & $2.51\pm0.35$ & $2.84\pm0.39$ \\
2 Dec. 2015  &  57.8 & $0.12\pm0.01$ & $0.36\pm0.01$ & $0.20\pm0.01$ & $3.03\pm0.18$ & $3.39\pm0.19$ \\ 
16 Mar. 2017 &  25.7 & $0.11\pm0.02$ & $0.34\pm0.02$ & $0.18\pm0.01$ & $3.18\pm0.10$ & $3.52\pm0.12$ \\
1 Jan. 2019  &   5.4 & $0.20\pm0.04$ & $0.50\pm0.09$ & $0.16\pm0.01$ & $2.97\pm0.18$ & $3.47\pm0.27$ \\
\hline
Average     &       & $0.14\pm0.03$ & $0.39\pm0.06$ & $0.20\pm0.02$ & $2.90\pm0.02$ & $3.29\pm0.23$ \\
\hline
\end{tabular}
\end{table}

\begin{figure}
\begin{center}
\includegraphics[width=\hsize]{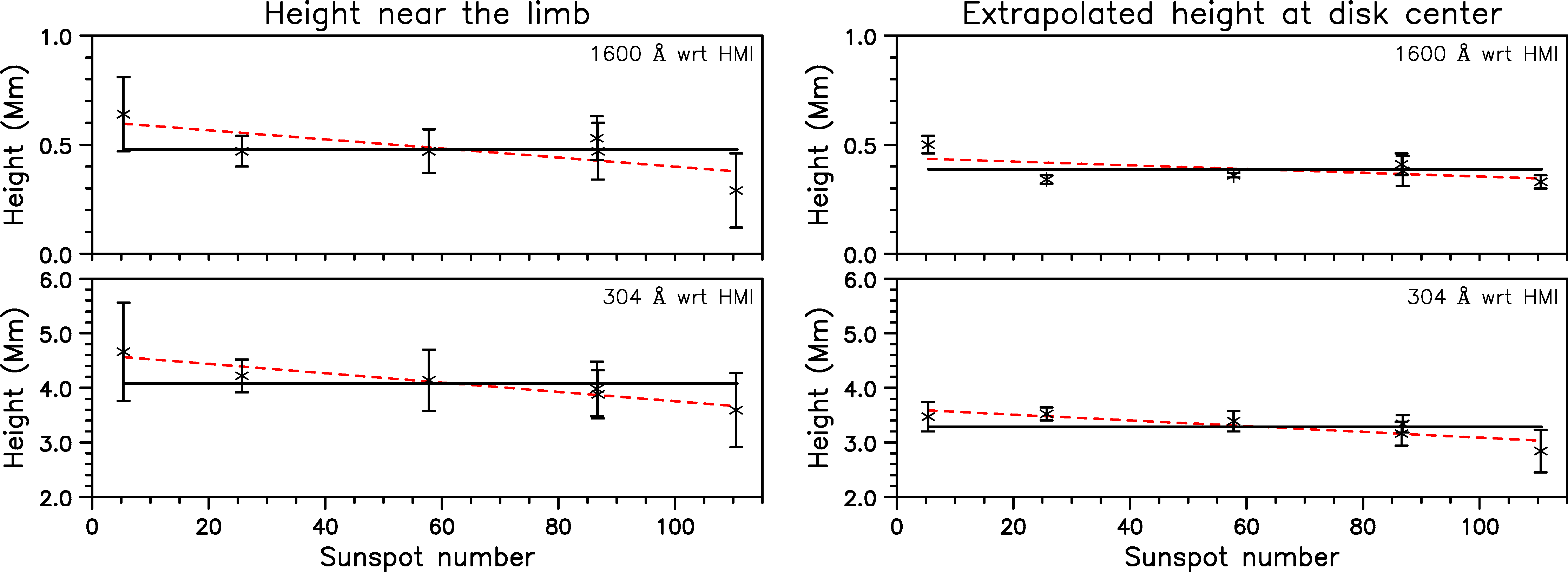}
\end{center}
\caption{The formation height near the limb and at disk center, as a function of the sunspot number, for the 1600 and 304\,\AA\ wavelength bands. {\it Black, full} lines are the average values, {\it red, dashed} lines the result of linear regression.}
\label{Fig:Sn}
\end{figure}

We note that in both Tables \ref{Table:heightsLimb} and \ref{Table:heightsCD} the maximum height was measured in 2019, close to the solar minimum, and the minimum height in 2014, close to the solar maximum. This is illustrated in Figure \ref{Fig:Sn}, where we plotted representative shifts as a function of the sunspot number. Although differences are small and within the error margins, this may disclose a solar cycle effect. However, for such a variation to be convincingly demonstrated, more data sets are required, and this is outside the scope of the present work. Moreover, this variation is not corroborated by our measurements of the solar radius, while \cite{2007A&A...476..369G} found no solar-cycle variation of the radius at 304 and 171\,\AA\ from EIT images. Still, \cite{2004A&A...420.1117S} reported a negative correlation between the solar radius at 17\,GHz and the sunspot number for the polar regions and a positive one for the full disk. Taking all of these into account, we consider that the evidence for a solar-cycle variation is weak.

\section{Summary and Conclusions}\label{concl}
We developed a method to measure the relative shift of disk features in two wavelength bands as a function of position angle and distance from the center of the solar disk and from that we computed the relative height of the network emission. We applied this method to six data sets, from 2012 to 2019, and we obtained the network height in the 1700, 1600, and 304\,\AA\ bands with respect to the level of the HMI emission at 6173\,\AA. As a proxy of the HMI network we used the absolute value of the longitudinal magnetic field. Near the limb, the apparent height of $|B_l|$ is 0.34\,Mm below the continuum network, probably due to the predominantly vertical orientation of the magnetic field; we corrected the limb heights for this effect.

\begin{figure}[h]
\begin{center}
\includegraphics[width=\hsize]{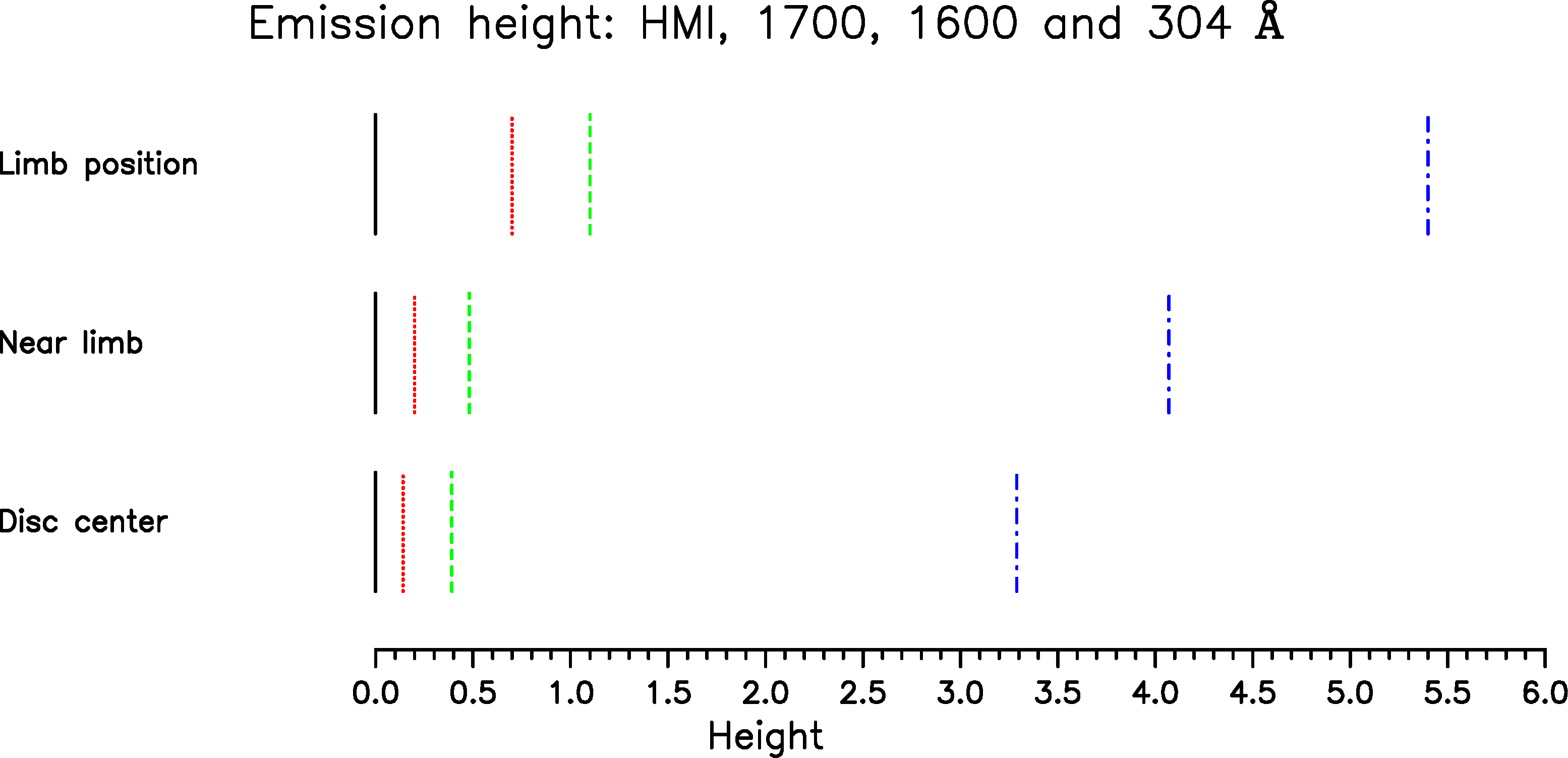}
\end{center}
\caption{The emission height in Mm with respect to HMI at disk center and near the limb, as well as the height of the limb, for the 1700, 1600, and 304\,\AA\ AIA wavelength bands ({\it red, full lines\/}, {\it green, dashed lines\/}, and {\it blue, dash-dotted lines} respectively).}
\label{Fig:Cartoon}
\end{figure}

No Pole--Equator differences were detected. We found a well-defined increase of the network height from the center of the disk to the limb. This is apparently due to opacity effects, as the radiation near the limb is formed higher than near the disk center. The lowest height is that of the 1700\,\AA\ network, just 0.14\,Mm above the HMI level at the disk center, followed by the 1600\,\AA\ network, 0.20\,Mm above the 1600 level; the corresponding values near the limb are 0.20\,Mm and 0.25\,Mm respectively. The average height of the 304\,\AA\ with respect to the 1600\,\AA\ emission ranges from 2.9\,Mm at disk center to 3.6\,Mm at the limb; thus the value of 3\,Mm reported by \cite{2018A&A...619L...6N} is between the two. Adding to that the height difference between the 1600\,\AA\ level and HMI, the 304\,\AA\ network height comes to 4.1 and 3.3\,Mm for the limb and disk center respectively. The formation heights as well as the limb height from \cite{2019SoPh..294...96A} are shown in Figure \ref{Fig:Cartoon}.

The deduced network heights near the limb are smaller than the heights of the limb at 1600 and 1700\,\AA\ and the height of 304\,\AA\ peak intensity reported by \cite{2019SoPh..294...96A}. As explained in the introduction, the limb height should be closer to the upper limit of the level of formation of the radiation rather than to the average; moreover, the limb height represents the entire atmosphere and not just the network that is measured here. Our values for the center of the disk for 1600 and 1700\,\AA\ (0.14 and 0.39\,Mm respectively) are not very far from the theoretical computations of \cite{2005ApJ...625..556F}, 0.36 and 0.43\,Mm, but their difference is (0.20\,Mm {\it vs.\/} 0.07\,Mm); we note, however, that the response of the 1600 and 1700\,\AA\ filters of TRACE, used by these authors, is not the same as that of the corresponding AIA filters. Thus, our results can help to check and refine atmospheric models.

Finally, our measurements show a slight increase of the network height from the maximum to the minimum of the solar cycle. However, the variation is comparable to our error estimates; thus we consider this finding as suggestive only. We believe that this issue warrants further investigation, with more data sets.

\begin{acks}
The author gratefully acknowledges use of data courtesy of NASA/SDO and the AIA and HMI science teams. He also thanks  A. Nindos and S. Patsourakos for comments on the manuscript and suggestions.
\end{acks}

\medskip\noindent{\footnotesize {\bf Disclosure of Potential Conflicts of Interest} The author declares that he has no conflicts of interest.}

\end{article} 
\end{document}